\documentstyle[12pt]{article}
\begin{document}

\def\subsubsection{\@startsection{subsubsection}{3}{\z@}{-3.25ex plus
 -1ex minus -.2ex}{1.5ex plus .2ex}{\large\sc}}
\renewcommand{\theequation}{\arabic{section}.\arabic{equation}}
\renewcommand{\thesection}{\arabic{section}}
\renewcommand{\thesubsection}{\arabic{section}.\arabic{subsection}}
\renewcommand{\thesubsubsection}
{\arabic{section}.\arabic{subsection}.\arabic{subsubsection}}
\pagestyle{plain}
\sloppy
\textwidth 150mm
\textheight 620pt
\topmargin 35pt
\headheight 0pt
\headsep 0pt
\topskip 1pt
\oddsidemargin 0mm
\evensidemargin 10mm
\setlength{\jot}{4mm}
\setlength{\abovedisplayskip}{7mm}
\setlength{\belowdisplayskip}{7mm}
\newcommand{\be}{\begin{equation}}
\newcommand{\ee}{\end{equation}}
\newcommand{\bea}{\begin{eqnarray}}
\newcommand{\ba}{\begin{array}}
\newcommand{\eea}{\end{eqnarray}}
\newcommand{\ea}{\end{array}}
\newcommand{\noin}{\noindent}
\newcommand{\ra}{\rightarrow}
\newcommand{\txs}{\textstyle}
\newcommand{\disp}{\displaystyle}
\newcommand{\scs}{\scriptstyle}
\newcommand{\scscs}{\scriptscriptstyle}
\newcommand{\sx}[1]{\sigma^{\, x}_{#1}}
\newcommand{\sy}[1]{\sigma^{\, y}_{#1}}
\newcommand{\sz}[1]{\sigma^{\, z}_{#1}}
\newcommand{\sP}[1]{\sigma^{\, +}_{#1}}
\newcommand{\sM}[1]{\sigma^{\, -}_{#1}}
\newcommand{\spm}[1]{\sigma^{\,\pm}_{#1}}
\newcommand{\Spm}{S^{\,\pm}}
\newcommand{\Sz}{S^{\, z}}
\newcommand{\hspf}{\hspace*{5mm}}
\newcommand{\hspt}{\hspace*{2mm}}
\newcommand{\vspf}{\vspace*{5mm}}
\newcommand{\vspt}{\vspace*{2mm}}
\newcommand{\hsix}{\hspace*{6mm}}
\newcommand{\hfour}{\hspace*{4mm}}
\newcommand{\vsix}{\vspace*{6mm}}
\newcommand{\vfour}{\vspace*{4mm}}
\newcommand{\vtwo}{\vspace*{2mm}}
\newcommand{\htwo}{\hspace*{2mm}}
\newcommand{\honecm}{\hspace*{1cm}}
\newcommand{\vonecm}{\vspace*{1cm}}
\newcommand{\htwocm}{\hspace*{2cm}}
\newcommand{\vtwocm}{\vspace*{2cm}}
\newcommand{\ru}{\rule[-2mm]{0mm}{8mm}}
\newcommand{\tinf}{\rightarrow \infty}
\newcommand{\udl}{\underline}
\newcommand{\ovl}{\overline}
\newcommand{\nwl}{\newline}
\newcommand{\nwp}{\newpage}
\newcommand{\clp}{\clearpage}
\newcommand{\simleq}{\raisebox{-1.0mm}
{\mbox{$\stackrel{\textstyle <}{\sim}$}}}
\newcommand{\simgeq}{\raisebox{-1.0mm}
{\mbox{$\stackrel{\textstyle >}{\sim}$}}}
\newcommand{\half}{\mbox{\small$\frac{1}{2}$}}
\newcommand{\smfrac}[2]{\mbox{\small$\frac{#1}{#2}$}}
\newcommand{\bra}[1]{\mbox{$\langle \, {#1}\, |$}}
\newcommand{\ket}[1]{\mbox{$| \, {#1}\, \rangle$}}
\newcommand{\exval}[1]{\mbox{$\langle \, {#1}\, \rangle$}}
\newcommand{\BIN}[2]
{\renewcommand{\arraystretch}{0.6}
\mbox{$\left(\ba{@{}c@{}}{\scs #1}\\{\scs #2}\ea\right)$}
\renewcommand{\arraystretch}{1}}
\newcommand{\UQSU}{\mbox{U$_{q}$[sl(2)]}}
\newcommand{\UQSUA}{\mbox{U$_{q}$[$\widehat{\mbox{sl(2)}}$]}}
\newcommand{\comm}[2]{[\, #1 , \, #2 \, ]}
\newcommand{\intpa}[1]{\mbox{\small $[ \, #1 \, ]$}}
\newcommand{\BC}{boundary conditions}
\newcommand{\TBC}{toroidal boundary conditions}
\newcommand{\ABC}{antiperiodic boundary conditions}
\newcommand{\PBC}{periodic boundary conditions}
\newcommand{\CBC}{cyclic boundary conditions}
\newcommand{\OBC}{open boundary conditions}
\newcommand{\XXZ}{XXZ Heisenberg chain}
\newcommand{\ZFQC}{Zamolodchikov Fateev quantum chain}
\newcommand{\FSS}{finite-size scaling}
\newcommand{\FSSS}{finite-size scaling spectra}
\newcommand{\DT}{duality transformation}
\newcommand{\POM}{Potts models}
\newcommand{\PM}{projection mechanism}
\newcommand{\PQC}{Potts quantum chain}
\newcommand{\TLA}{Temperley-Lieb algebra}
\newcommand{\PTLA}{periodic Temperley-Lieb algebra}
\newcommand{\HA}{Hecke algebra}
\newcommand{\VA}{Virasoro algebra}
\newcommand{\CFT}{conformal field theory}
\newcommand{\SS}{steady state}
\newcommand{\AI}{${\rm A}_{\rm I}$ }
\newcommand{\AII}{${\rm A}_{\rm II}$ }
\newcommand{\BI}{${\rm B}_{\rm I}$ }
\newcommand{\BII}{${\rm B}_{\rm II}$ }

\begin{titlepage}
\thispagestyle{empty}
\begin{center}
\vspace*{1cm}
{\bf \Large Phase transitions in an exactly soluble one-dimensional
exclusion process}\\[28mm]

{\Large {\sc
G. Sch\mbox{\"u}tz\footnotemark[1] \vtwo \\
and \vfour\\ E. Domany}\footnotemark[2]
} \\[5mm]

\begin{minipage}[t]{13cm}
\begin{center}
{\small\sl
  \footnotemark[1]
  Department of Physics,
                Weizmann Institute, \\
  Rehovot 76100,
  Israel
  \\[2mm]
  \footnotemark[2]
  Department of Electronics,
                Weizmann Institute, \\
  Rehovot 76100,
  Israel}
\end{center}
\end{minipage}
\vspace{28mm}
\end{center}
{\small
We consider an exclusion process, with particles injected with rate
$\alpha$ at
the origin and removed with rate $\beta$
at the right boundary of a one-dimensional
chain of sites. The particles are allowed to hop onto unoccupied
sites, to the right only. For the special case of $\alpha=\beta=1$
the model was solved previously by Derrida et al. Here we extend the
solution to general $\alpha,\beta$. The phase diagram obtained from
our exact solution differs from the one predicted by the mean field
approximation.
}
\\
\vspace{5mm}\\
\udl{Key words:} asymmetric exclusion process,
steady state, boundary induced phase transitions
\end{titlepage}

\newpage
\baselineskip 0.3in
\section{Introduction}
\setcounter{equation}{0}

One-dimensional
asymmetric exclusion models  \cite{1}
are of interest for various reasons. They are closely
related to vertex models \cite{KDN},
growth models \cite{KS} and, in the continuum limit,
the KPZ equation \cite{KPZ} and the noisy Burgers's equation.

Various types of phase transitions occur as a consequence of the
interplay of particle transport with a localized defect or
inhomogenity.
Suitably chosen boundary conditions can represent the effect of such
a defect in an otherwise homogenous system. Such transitions have
been the focus of many recent studies
\cite{1},\cite{K}-\cite{G2}. Some of these models could be solved exactly
and allow for a detailed study of their steady state properties
such as the density profile or density correlations \cite{1,G1,DE,G2}.

Totally asymmetric simple-exclusion models with nearest neighbour hopping
can be divided into four classes according to the dynamics (sequential
or parallel) and the choice of \BC\ (open or periodic). In all
these models each lattice site $i$ in a chain of $N$ sites
is either occupied by a single particle ($\tau_i = 1$)
or empty ($\tau_i = 0$)
and a particle can hop to the neighbouring site in one direction if
this site is empty.\footnote{A model with two different kinds of
particles and nearest neighbour hopping has been studied in
\protect\cite{KM}, a simple-exclusion model where particles can
hop over several lattice sites in each time step is discussed in
\protect\cite{NS}.} By convention, we choose the direction of hopping
as to the right. The dynamics can be chosen either sequential
as in refs. \cite{1,K,JL,DE} or parallel \cite{KDN,G1,G2}.
In the case of sequential dynamics which we study in this paper
particles jump independently and randomly in each time step
according to the following rules: At each time step $t \ra t+1$
one chooses at random one pair of sites ($i,i+1$) with $1\leq i
\leq N-1$. If there is a particle
on site $i$ and site $i+1$ is empty, then the particle will jump
from $i$ to $i+1$. All other configurations do not change, i.e.,
\be\label{1}\ba{rcl}
\tau_i(t+1) & = & \tau_i(t) \tau_{i+1}(t) \vfour \\
\tau_{i+1}(t+1) & = & \tau_{i+1}(t) + (1-\tau_{i+1}(t))\tau_i(t) \htwo .
\ea\ee
In the case of parallel updating the lattice is divided into
neighbouring pairs of sites and some stochastic
hopping rules are applied in
parallel to each pair in a first half time step. In the second
half time step the pairs are shifted by one lattice unit and the
same rules are applied again \cite{KDN,G1,G2}. Both models
can be defined with either
periodic or \OBC . When periodic \BC\ are used, a non-trivial phase
diagram can be observed by introducing, for example,
a single defect \cite{JL,G1}. With \OBC\ particles are
injected with  rate $\alpha$ at the
left boundary (which we shall call the {\it origin}), and absorbed
with  rate $\beta$ at the right {\it boundary}
\cite{1,K,DE,G2}. Injection and absorption are implemented in the
following way: when
one considers the pair (0,1) where site 0 represents the origin,
then the occupation number $\tau_1(t+1)$ of site 1 at time $t+1$
is given by
\be\label{2}\ba{rcl}
\tau_1(t+1) & = & 1 \htwo \mbox{with probability $\tau_1(t) +
                    \alpha (1-\tau_1(t))$} \vfour \\
\tau_1(t+1) & = & 0 \htwo \mbox{with probability $(1-\alpha)
                           (1-\tau_1(t))$} \htwo .
\ea\ee
On the other hand, considering the pair $(N,N+1)$ where site
$N+1$ represents the (right) boundary, the occupation number
$\tau_N(t+1)$ at site $N$ after one time step is
\be\label{3}\ba{rcl}
\tau_N(t+1) & = & 1 \htwo \mbox{with probability $(1-\beta)\tau_N(t)$}
                    \vfour \\
\tau_N(t+1) & = & 0 \htwo \mbox{with probability $1 - (1-\beta)
                          \tau_N(t)$} \htwo .
\ea\ee
The model defined by
eqs. (\ref{1}) - (\ref{3}) can be viewed as
a homogeneous system connected to a reservoir of fixed particle
density $\alpha$ at the origin and fixed density $1-\beta$ at the
boundary.

The model with parallel dynamics and \PBC\ with a defect \cite{G1}
was solved by the Bethe
ansatz.
Recently some \SS\ properties
of the model with parallel dynamics and \OBC\  were also
found \cite{G2}.
The Bethe ansatz was used also to solve the model with sequential
updating (\ref{1}) and translationally invariant \PBC\ without defect
\cite{GS}.
The case of sequential dynamics with \OBC\ was studied by Krug \cite{K}
and by Derrida et al \cite{1}.

Krug studied numerically the steady state behavior of this
model on the line $\beta=1$
(Fig.~1). He  found, at $\alpha=1/2$, a phase
transition and an associated diverging length
scale \cite{K}. For $\alpha< 1/2$ he found
an exponential decay of the profile to its bulk value with increasing
distance $r$ from the boundary, while for $\alpha>1/2$ the profile
decayed as $r^{-1/2}$. Derrida et al
\cite{1} expressed the exact \SS\ and the \SS\
density distribution
for arbitrary $\alpha$ and $\beta$ in terms of
recursion relations (eqs. (\ref{12}) - (\ref{14}) below). These
recursions were solved explicitly only for $\alpha=\beta=1$.
For this case they showed that the
density profile approaches algebraically its bulk value $\rho_{bulk}
=1/2$ with increasing distance $x$ from the origin,
$\rho-1/2 \sim x^{-1/2}$. The same behavior characterizes the
approach of
$\rho_{bulk}$ from below as $r^{-1/2}$, with increasing distance $r$
from  the boundary, confirming the numerical result of Krug.

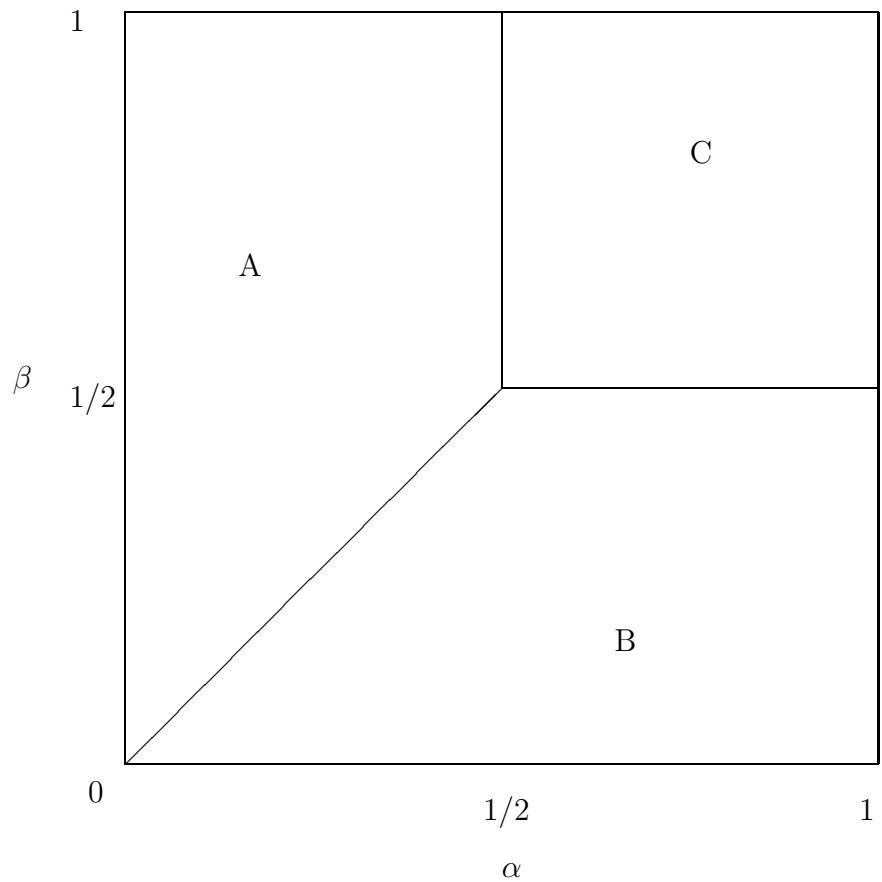
\begin{figure}
\setlength{\unitlength}{2.5mm}
\begin{center}
\begin{picture}(50,50)
\put (2,2){0}
\put (1,23){1/2}
\put (1,43){1}
\put (23,1){1/2}
\put (43,1){1}
\put (24,-2){$\alpha$}
\put (-2,24){$\beta$}
\put (4,4){\framebox(40,40)}
\put (10,30){A}
\put (34,36){C}
\put (30,10){B}
\put (24,24){\line(1,0){20}}
\put (24,24){\line(0,1){20}}
\put (4,4){\line(1,1){20}}
\end{picture}
\end{center}
\caption{\protect\small
Mean field phase diagram of the model in the $\alpha-\beta$ plane
as obtained in \protect\cite{1}. Region A is the low density phase,
region B  the high density phase and region C is the maximal current
phase. The phases are separated by the curves $\alpha=\beta<1/2$
and $\alpha=1/2$, $\beta>1/2$ and $\beta=1/2$, $\alpha>1/2$ respectively.
The sign of the slope of the density profile (as shown in the insets)
changes when the line $\alpha=1-\beta$ is crossed. Note that this is
{\em not} a phase transition line.}
\end{figure}

The phase diagram in the whole $\alpha-\beta$ plane
was obtained in \cite{1}
by a {\it mean field} calculation. Three phases were identified (Fig.~1).
In a low density phase $A$, found
for $\alpha < \beta$ and $\alpha < 1/2$,
the density profile approaches $\rho_{bulk}=\alpha$ exponentially
with $r$. This supports Krug's
observation of an exponential behaviour for $\beta=1$, $\alpha<1/2$.
A high density phase $B$ was found for $\alpha > \beta$, $\beta<1/2$,
which is related to the low
density phase by a particle-hole symmetry. Here the profile approaches
$\rho_{bulk}=1-\beta$ exponentially with $x$, the distance from the
origin.
Finally, for $\alpha,
\beta > 1/2$ the system is in the maximal current phase $C$. In this
phase mean field predicts a power law for the profile, with exponent
$\kappa=1$, whereas the exact result \cite{1} yields $\kappa=1/2$.

Here we present the exact solution to the recursion relations
giving the steady state and the density profile for arbitrary
values for $\alpha$ and $\beta$. We show that the phase diagram
has a richer structure than that
predicted by  mean field.
In particular, we show that there is a phase transition with an
associated diverging length scale along the two lines,
$\alpha=1/2$
and $\beta=1/2$, dividing both the low density phase and the high density
phase found in the mean field calculation into two different phases.
In the low density phase \AI defined by $\alpha < \beta < 1/2$
the profile is exponential, as predicted by the mean field
calculation. The situation is different, however,
in the low density phase \AII , defined by $\alpha < 1/2$
and $\beta > 1/2$; there the profile approaches $\rho_{bulk}=\alpha$
as $r^{-3/2}\exp{(-r/\xi)}$ for $r \gg 1$. This was expected neither
from the mean field approach nor from the numerical results of Krug.

The paper is organized as follows. In sec.~2 we present the
recursion relations obtained in \cite{1} and give an exact
solution for arbitrary $\alpha$ and $\beta$. In sec.~3 we draw
some conclusions from these results and derive the exact phase
diagram. Then we study the density profile
in the various phases for large systems (sec.~4) and in sec.~5
discuss in detail the various phase transitions that were identified.

\section{Exact solution of the recursion relations}
\setcounter{equation}{0}

The \SS\ of the model defined in eqs. (1) - (3) is given in terms of
the
quantities $P_N(\tau_1,\tau_2,\dots,\tau_N)$ which are the probabilities
of finding the specific configuration of particles represented by
the occupation numbers $(\tau_1,\tau_2,\dots,\tau_N)$ on the chain
with $N$ sites. It turns out to be more convenient to work with
unnormalized probabilities  $f_N(\tau_1,\tau_2,\dots,\tau_N)$
related to $P_N(\tau_1,\tau_2,\dots,\tau_N)$ by
\be\label{4}
P_N(\tau_1,\tau_2,\dots,\tau_N) = f_N(\tau_1,\tau_2,\dots,\tau_N)/Z_N
\ee
where
\be\label{5}
Z_N = \sum_{\tau_1=0,1} \dots \sum_{\tau_N=0,1}
f_N(\tau_1,\tau_2,\dots,\tau_N) \htwo .
\ee
As shown in \cite{1} all the $f_N(\tau_1,\tau_2,\dots,\tau_N)$
can be obtained recursively from the corresponding quantities
in a system with $N-1$ sites (eqs. (8) and (9) of ref. \cite{1}).
Here we are interested only in the average occupation number
$\langle \tau_i \rangle_N$ of site $i$ in a system of length $N$,
given by
\be\label{6}
\langle \tau_i \rangle = T_{N,i} / Z_N
\ee
with
\be\label{7}
T_{N,i} = \sum_{\tau_1=0,1} \dots \sum_{\tau_N=0,1} \tau_i
f_N(\tau_1,\tau_2,\dots,\tau_N) \htwo .
\ee

The normalization $Z_N$ and the unnormalized particle density
$T_{N,i}$ can be computed from the quantities
\be\label{8}
Y_{N,K} = \sum_{\tau_1=0,1} \dots \sum_{\tau_N=0,1} (1-\tau_N)
          (1-\tau_{N-1})\dots(1-\tau_K)
          f_N(\tau_1,\tau_2,\dots,\tau_N)
\ee
and
\be\label{9}
X_{N,K}^p = \sum_{\tau_1=0,1} \dots \sum_{\tau_N=0,1} (1-\tau_N)
            \dots(1-\tau_K)\tau_p
          f_N(\tau_1,\tau_2,\dots,\tau_N)
\ee
by defining
\be\label{10}
Y_{N,N+1} = Z_N
\ee
and
\be\label{11}
X_{N,N+1}^p = T_{N,p} \htwo .
\ee
(Note that we made a slight change in notation as compared to
ref.~\cite{1}. There the quantities $Y_{N,K}$ were denoted $Y_N(K)$
and the quantities $X_{N,K}^p$ were denoted $X_N(K,p)$.)

The reason for the introduction of $Y_{N,K}$ for $1 \leq K \leq N+1$
and $X_{N,K}^p$ for $p+1 \leq K \leq N+1$ is that they can be obtained
from the following closed recursions \cite{1}
\be\label{12}
\ba{rcl}
Y_{N,1} & = & \beta Y_{N-1,1} \vfour \\
Y_{N,K} & = & Y_{N,K-1} + \alpha \beta Y_{N-1,K} \hfour
              \mbox{for $2 \leq K \leq N$} \vfour \\
Y_{N,N+1} & = & Y_{N,N} + \alpha Y_{N-1,N}
\ea\ee
with the initial condition
\be\label{13}
\ba{rcl}
Y_{1,1} & = & \beta \vfour \\
Y_{1,2} & = & \alpha+\beta \htwo .
\ea\ee
These recursions can be simplified somewhat by
extending the range of definition of $K$ to $1 \leq K \leq N+2$.
If we set $K=N+1$ in the second of eq. (\ref{12}), the resulting
equation is precisely the third of (\ref{12}), provided we use
the extended definition
$Y_{N-1,N+1} = \beta^{-1} Y_{N-1,N}$ for $\beta\neq 0$.
Similarly,
eqs. (\ref{13}) become a consequence of (\ref{12}) by redefining the
initial condition as $Y_{0,1}=1$. These extensions of the definitions
of the quantities $Y_{N,K}$ are useful in some of the
calculations presented below.

Once the $Y_{N,K}$ are determined, the $X_{N,K}^p$ can be obtained
from the recursion relations \cite{1}
\be\label{14}
\ba{rcll}
X_{N,K}^p & = & X_{N,K-1}^p + \alpha \beta X_{N-1,K}^p \hfour &
              \mbox{for $p+2 \leq K \leq N$} \vfour \\
X_{N,N+1}^p & = & X_{N,N}^p + \alpha X_{N-1,N}^p &
              \mbox{for $1 \leq p \leq N-1$}
\ea\ee
with the initial condition
\be\label{15}
X_{N,p+1}^p = \alpha \beta Y_{N-1,p+1}  \hfour
              \mbox{for $1 \leq p \leq N$}
\ee
where we used the extended definitions $Y_{0,1}$ and $Y_{N,N+2}$
of the $Y_{N,K}$. Solving these recursion relations gives the exact
average occupation numbers $\langle \tau_i \rangle$ through eqs.
(\ref{6}), (\ref{10}) and (\ref{11}). This was done in ref. \cite{1}
for $\alpha=\beta=1$. Here we present the solution for arbitrary
$\alpha$ and $\beta$.

For a solution of these recursion relations and
initial conditions define the functions $G_{N,K}^{M}(x)$ by
\be\label{16}
G_{N,K}^{M}(x) = \sum_{r=0}^{M-1} b_{N,K}(r) x^r \hspace{2cm} (N\geq 1)
\ee
with
\be
b_{N,K}(r) = \left( \ba{c} K-2+r \\ K-2 \ea \right) -
             \left( \ba{c} K-2+r \\  N  \ea \right) \hspace{2mm} .
\ee
For later
convenience also define $b_{0,1}(0)=G_{0,0}^1=G_{0,1}^1
=1$. As a result of the symmetries of the coefficients $b_{N,K}(r)$
these functions satisfy various relations given in the appendix.
In particular, from the recursion relations (\ref{A1}) and the
special values (\ref{A2}) one can show that the quantity
\be\label{17}
Y_{N,K}(\alpha,\beta) = \beta^N G_{N,K}^N(\alpha) + \sum_{s=0}^{K-2}
\alpha^{N-s} \beta^{N-K+1+s} G_{N,N}^{s+1} (\alpha)
\ee
solves the recursion relations eq. (\ref{12}) with the initial
conditions (\ref{13}).

Relations (\ref{14}) with initial condition (\ref{15}) are satisfied by
\be\label{18}
\ba{rcl}
X_{N,K}^p(\alpha,\beta) & = & \disp \sum_{r=0}^{N-p} b_{N-p,K-p}(r)
\alpha^{r+1} \beta^{r+1} Y_{N-r-1,p+1}(\alpha,\beta) \; + \vfour \\
 &   & \disp \beta^{N-K+2} \sum_{r=0}^{K-p-2} \alpha^{N+1-p-r}
G_{N-p,N-p}^{K-p-1-r}(\beta) Y_{p-1+r,p+1}(\alpha,\beta) \hspace{2mm} .
\ea\ee

Using the first of equations (\ref{A2}),
one obtains from this
\be
\label{19}
Z_N = Y_{N,N+1} = \sum_{s=0}^{N} \beta^s \alpha^{N-s}
G_{N,N}^{s+1}(\alpha)
\ee
and after some computation, involving relabeling of indices, we get
\be
\label{20}
T_{N,p} = X_{N,N+1}^p = \alpha \beta \sum_{s=0}^{N-p}
\alpha^s G_{N-p,N-p}^{s+1}(\beta) Y_{N-s-1,p+1}(\alpha,\beta) \htwo .
\ee
This expression is exact for any $N \geq 1$, $1 \leq p \leq N$.
Substitution in (2.3) gives the exact density profile for
arbitrary $\alpha$ and $\beta$.

Eqs. (\ref{19}) and (\ref{20}) provide also an exact expression
for the conserved current
$j=\langle \tau_i \rangle - \langle \tau_i \tau_{i+1} \rangle = const.$,
and consequently, for the correlation function
$\langle \tau_i \tau_{i+1} \rangle$. To see this, note that
taking $i=N$ one obtains
$\langle \tau_N \tau_{N+1} \rangle = (1-\beta) \langle \tau_N \rangle$
since site $N+1$ represents the reservoir of constant density
$1-\beta$. Therefore one has \cite{1}
\be\label{20a}
j = \beta \langle \tau_N \rangle \htwo .
\ee
On the other hand, taking $i=0$ one gets
$\langle \tau_0 \tau_{1} \rangle = \alpha \langle \tau_1 \rangle$
since site 0 is the reservoir of constant density $\alpha$. Thus
we also have \cite{1}
\be\label{20b}
j = \alpha (1 - \langle \tau_1 \rangle) \htwo .
\ee
Since, however, our exact result yields $\langle \tau_i \rangle$ for
any $i$, we can calculate the exact current $j$, and hence
$\langle \tau_i \tau_{i+1} \rangle$.

\section{Discussion of the density profile}
\setcounter{equation}{0}

In order to analyse the density,
it is convenient to study the quantity
\be
t_N(p) = ( T_{N,p+1} - T_{N,p} ) / Z_N
\ee
which becomes the spatial derivative
of the density profile
in the continuum limit.
It turns out to be given by ($p\neq N$)
\be
\label{21}
t_N(p) = (1-\alpha - \beta) \beta^p G_{N-p,N-p}^{N-p}(\beta)
\alpha^{N-p} G_{p,p}^p (\alpha) / Z_N \hspace{2mm} .
\ee
which for $\alpha \neq 1-\beta$ can be more conveniently
written in the form
\be\label{22}
t_N(p) =
F_p(\alpha) F_{N-p}(\beta) / \tilde{Z}_N
\ee
with
\be\label{23}
F_N(x) = x^{-N-1} \, G_{N,N}^N(x)
\ee
and
\be\label{25}
\tilde{Z}_N  = \frac{Z_N}{(1-\alpha-\beta)\alpha^{N+1}\beta^{N+1}} =
\left\{ \ba{ll} \disp
\frac{F_N(\beta) - F_N(\alpha)}{\alpha(1-\alpha) - \beta(1-\beta)}
\hsix & \alpha \neq \beta,1-\beta \vfour \\
\disp - \frac{F_N'(\beta)}{1-2\beta} & \alpha = \beta \neq \half
 \ea \right.
\ee
where the prime denotes the derivative w.r.t. $\beta$. The last of the
two eq. (\ref{25}) can be obtained by changing
the order of summation in (\ref{19}).

For $\alpha=1-\beta$ one obtains
directly from (\ref{21}) that $t_N(p) = 0$, i.e., the profile is
constant on this curve. This result was already obtained in \cite{1}.
{}From (\ref{22}) we learn that up to the amplitude $\tilde{Z}$,
the derivative $t_N(p)$ of
the density profile can be written as a
product of two functions; one  of $\alpha$ and the other of
$\beta$: $t_N(p)
\propto F_p(\alpha) F_{N-p}(\beta)$.

This fact has important and surprising consequences. It clearly implies
that phase transitions (i.e. non-analytic changes in the $p$-dependence
of the density profile) can occur on two kinds of lines:
$\alpha=\alpha_c$ and $any$ $\beta$, or $\beta=\beta_c$ and {\it any}
$\alpha$. Hence if a phase transition is predicted to occur on the
$\beta>1/2$ segment of the line $\alpha=1/2$ (the mean field transition
to the maximal current phase), then the transition {\em must} extend to
the $\beta < 1/2$ regime as well! This means that instead of a single
high density phase $B$, predicted by mean field, there  are, in fact,
two such phases. Indeed, analysis of the function $F_p(x)$, presented
below, reveals that its dependence on $p$ changes at $x=1/2$. Similar
considerations hold for the
line $\beta=1/2$ and $\alpha<1/2$, which separates the low density
phase $A$ into two distinct phases (Fig.~2). These new transitions
were not found by the mean field calculation.

\begin{figure}
\setlength{\unitlength}{2.5mm}
\begin{center}
\begin{picture}(50,50)
\put (2,2){0}
\put (1,23){1/2}
\put (1,43){1}
\put (23,1){1/2}
\put (43,1){1}
\put (24,-2){$\alpha$}
\put (-2,24){$\beta$}
\put (4,4){\framebox(40,40)}
\put (11,33){${\rm A}_{\rm II}$}
\put (33,11){${\rm B}_{\rm II}$}
\put (17,10){${\rm B}_{\rm I}$}
\put (10,17){${\rm A}_{\rm I}$}
\put (33,33){C}
\put (4,24){\line(1,0){40}}
\put (24,4){\line(0,1){40}}
\put (4,4){\line(1,1){20}}
\end{picture}
\end{center}
\caption{\protect\small
Exact phase diagram of the model in the $\alpha-\beta$ plane.
The low (high) density phase shown in Fig.~1 is divided into two
phases ${\rm A}_{\rm I}$ and ${\rm A}_{\rm II}$ (${\rm B}_{\rm I}$ and
${\rm B}_{\rm II}$) along the curve $\beta=1/2$
($\alpha=1/2$).}
\end{figure}
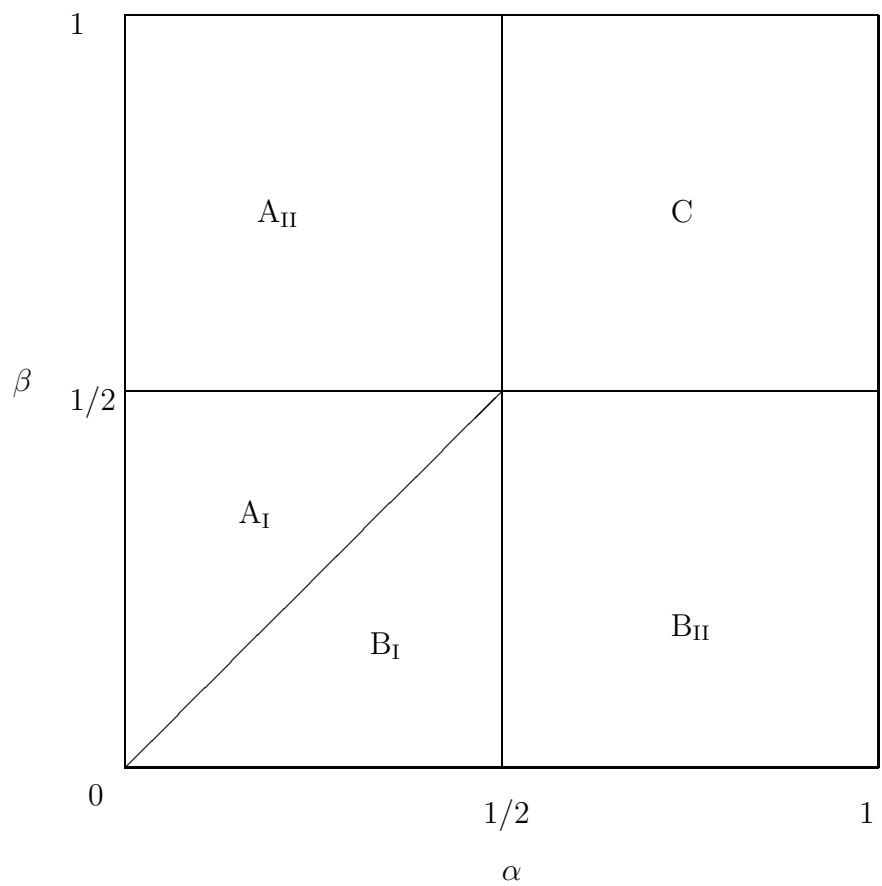

Another unexpected consequence of the separability into a product
is the existence
of {\em two} independent
length scales in the model, one determined by the
injection rate $\alpha$, the other one by the absorption rate $\beta$.
This is surprising, as one might believe
that only the larger of these two quantities determines the
behaviour of the system. In fact, as long as the system is not in
the maximal current phase, this indeed is the case as far as the
current $j=\langle \tau_i \rangle - \langle \tau_i \tau_{i+1} \rangle$
is concerned: In the continuum limit one has
$j=\beta(1-\beta)$ for $\alpha > \beta$, $\beta < 1/2$, and
$j=\alpha(1-\alpha)$ for $\beta > \alpha$, $\alpha < 1/2$, (whereas
$j=1/4$ if both $\alpha$ and $\beta$ are larger than 1/2). Since
in the mean field calculation the
shape of the density profile is determined by only the current,
phase transitions are seen neither at $\alpha = 1/2$, $\beta
< 1/2$, nor at $\beta = 1/2$, $\alpha < 1/2$.
Prior to presenting an explanation for the unexpected
existence of the  additional
phases and phase transitions, we study the density profile
in the thermodynamic limit $N\rightarrow\infty$.

\section{Density profile in the large $N$ limit}
\setcounter{equation}{0}

We want to discuss the density profile of a large system ($N\gg 1$)
as a function of the space coordinate $p$,
at large distances from both ends, i.e., we consider $p \gg 1$
and $r=N-p \gg 1$. So we need an asymptotic expression for $F_L(x)$
for large $L$. Splitting $F_L(x)$ into two pieces $F_L^{(1)}(x)$ and
$F_L^{(2)}(x)$ as in
eq. (\ref{A9}) allows for an expansion in $1/L$. For $x<1/2$ the
dominating contribution is $F_L^{(1)}$ since $F_L^{(2)}/F_L^{(1)}
\propto \exp{(-aL)}$ with some constant $a$:
\be\label{26}
F_L(x) = \frac{1-2x}{\left(x(1-x)\right)^{L+1}}
\left( 1 + O(e^{-aL})\right) \htwocm x < \half \htwo .
\ee
If $x> 1/2$ then $F_L^{(1)}=0$ and up to order $1/L$
\be\label{27}
\ba{rcl}
F_L(x) & = & \disp
 \frac{c_L}{(1-2x)^2} \left( 1 + O(L^{-1}) \right) \vfour \\
 & = & \disp \frac{4^L}{ (1-2x)^2 \sqrt{\pi} L^{3/2} }
 \left( 1+ O(L^{-1}) \right) \htwocm x > \half \htwo .
\ea\ee
This expression diverges for $x \ra 1/2$.
For $x=1/2$ one obtains (see (\ref{A11}))
\be\label{28}
\ba{rcl}
F_L(x) & = & \disp
2 \frac{4^L}{\sqrt{\pi L}} \left( 1 + O(L^{-1}) \right) \htwo .
\ea\ee

Using the expansions eqs. (\ref{26}) - (\ref{28}) and the expression
(\ref{25}) for the normalization $\tilde{Z}_N$
one can compute the shape of the density profile given by $t_N(p)$.
We define a length scale $\xi_{\sigma}$ by
\be
\xi_{\sigma}^{-1} = -\log{(4\sigma(1-\sigma))}.
\ee
As $\sigma$ reaches 1/2, $\xi_{\sigma}$ diverges.
For the various phases \AI - C (Fig.~2)
one finds in the large $N$ limit
(such that $1 \ll p$, $1\ll N-p$, i.e., $p$ is far from both
ends of the system) the following results:
\vspace*{1cm} \\
\udl{\bf High density phase \BI:}
\vspace*{1cm} \\
This phase is defined by the region $\beta < \alpha < 1/2$.
{}From the expansion (\ref{26}) one finds an
exponential decay of the density profile with a
exponential decay with
length scale $\xi^{-1} = \xi_{\alpha}^{-1}
 - \xi_{\beta}^{-1}$,
\be\label{31}
\ba{rcl}
t_N(p) & = & \disp  (1-2\alpha) \left(
1 - \frac{4\beta(1-\beta)}{4\alpha(1-\alpha)} \right)
\left( \frac{4\beta(1-\beta)}{4\alpha(1-\alpha)} \right)^{p} \vfour \\
       & = & \disp  (1-2\alpha) \left(
1 - \frac{4\beta(1-\beta)}{4\alpha(1-\alpha)} \right)
\mbox{e}^{-p/\xi} \htwo .
\ea\ee
The density approaches its bulk value $\rho_{bulk} = 1- \beta$
from below. One has $j=\beta(1-\beta)$ and from (\ref{20b})
$\langle \tau_1 \rangle = 1 - \beta(1-\beta)/\alpha <1-\beta=\langle
\tau_N \rangle$.
\vonecm \\
\udl{\bf Transition line from high density phase \BI
to high density phase \BII:}
\vspace*{1cm} \\
On approaching $\alpha=1/2$ from below in the region $\beta<1/2$, we
find that
$\xi_{\alpha}$ diverges but $\xi_{\beta}$ remains finite. For
$\alpha=1/2$ the slope of the profile is of the form
\be\label{31a}
t_N(p) \sim p^{-z_{\alpha}} \mbox{e}^{-p/\xi_{\beta}} \htwo .
\ee
The values of the length scale $\xi_{\beta}$ and the exponent
$z_{\alpha}$ can be read off the exact expression (\ref{28}) which
for large $N$ becomes
\be\label{32}
\ba{rcl}
t_N(p) & = & \disp  \frac{(1-2\beta)^2}{2 \sqrt{\pi}}
 \frac{(4\beta(1-\beta))^p}{p^{1/2}} \vfour \\
       & = & \disp  \frac{(1-2\beta)^2}{2 \sqrt{\pi}}
             p^{-1/2} \mbox{e}^{-p/\xi_{\beta}} \htwo .
\ea\ee
The current and the boundary values are given by the same expressions
as in the high density phase I.
\vspace*{1cm} \\
\udl{\bf High density phase \BII:}
\vspace*{1cm} \\
On crossing the phase transition line into the high density phase \BII
defined by $\alpha>1/2$ and $\beta<1/2$,
the decay exponent changes to $z_{\alpha}= 3/2$ (see eq. (\ref{27}))
and one obtains
\be\label{33}
\ba{rcl}
t_N(p) & = & \disp  \frac{(1- \alpha -\beta)(\alpha - \beta)}
{(1-2\alpha)^2 \sqrt{\pi}}
 \frac{(4\beta(1-\beta))^p}{p^{3/2}} \vfour \\
       & = & \disp  \frac{(1- \alpha -\beta)(\alpha - \beta)}
{(1-2\alpha)^2 \sqrt{\pi}}
             p^{-3/2} \mbox{e}^{-p/\xi_{\beta}} \htwo .
\ea\ee
The current and the boundary values are given by the same expressions
as in the high density phase I, but note that the slope of the profile
changes sign on the curve $\alpha=1-\beta$. Along this curve the
density is constant, $\langle \tau_i \rangle = \rho_{bulk} = 1- \beta$
for $1 \leq i \leq N$. For $\alpha > 1- \beta$ the slope is negative.
\vspace*{1cm} \\
\udl{\bf Transition from high density phase \BII to
the maximal current phase C:}
\vspace*{1cm} \\
When $\beta$ reaches the critical value 1/2 in the region
$\alpha > 1/2$, $\beta \leq 1/2$, then also
$\xi_{\beta}$ diverges and the slope of the profile is given by
\be\label{34}
t_N(p) = -\frac{1}{4 \sqrt{\pi}} (1-\frac{p}{N})^{-1/2} p^{-3/2} \htwo .
\ee
Near the origin ($1 \ll p \ll N$) we can neglect the piece with
$p/N$ in (\ref{34}), so the slope is dominated by $p^{-z_{\alpha}}$ with
$z_{\alpha} = 3/2$. In the boundary region ($p=N-r$, $1 \ll r \ll N$)
the shape of the profile is determined by $r^{-z_{\beta}}$ with
$z_{\beta} = 1/2$, but
the amplitude of $t_N(r)$ is only of order $1/N$. Therefore,
up to corrections of order $1/N$, the profile near the boundary
is flat, whereas it decays as $p^{-1/2}$ with the distance $p$
from the origin to its bulk value $\rho_{bulk} = 1/2$. The
The current reaches its maximal value $j=1/4$ and one finds
$\langle \tau_N \rangle = \rho_{bulk} = 1/2$ and $\langle \tau_1
\rangle = 1 - 1/(4\alpha)$.
\vspace*{1cm} \\
\udl{\bf Maximal current phase:}
\vspace*{1cm} \\
If $\beta > 1/2$ and $ \alpha > 1/2$, the derivative
$t_N(p)$ depends neither on $\alpha$ nor on $\beta$. Near
the origin and near $N$ the slope of the profile is determined by
$z_{\alpha}=z_{\beta}=3/2$:
\be\label{35}
t_N(p) = -\frac{1}{4 \sqrt{\pi}} (1-p/N)^{-3/2} p^{-3/2}
\ee
Therefore the density approaches its bulk value $\rho_{bulk} = 1/2$
as $p^{-1/2}$ with the distance $p$ from the origin from above
and as $r^{-1/2}$ with the distance $r=N-p$ from the boundary from
below. The current takes its maximal value $j_{max}=1/4$ throughout the
phase and one obtains $\langle \tau_N \rangle = 1/(4\beta)$ and
$\langle \tau_1 \rangle = 1 - 1/(4\alpha)$.
\vspace*{1cm} \\
\udl{\bf Low density phase \AI:}
\vspace*{1cm} \\
This phase is defined by $\alpha < \beta < 1/2$ and
is related to the high density phase \BI by a particle-hole
symmetry and therefore the decay is exponential. One finds
\be\label{29}
\ba{rcl}
t_N(p) & = & \disp (1-2\beta) \left(
1 - \frac{4\alpha(1-\alpha)}{4\beta(1-\beta)} \right)
\left( \frac{4\alpha(1-\alpha)}{4\beta(1-\beta)} \right)^{N-p} \vfour \\
       & = & \disp (1-2\beta) \left(
1 - \frac{4\alpha(1-\alpha)}{4\beta(1-\beta)} \right) \mbox{e}^{-r/\xi}
\ea\ee
with a length scale $\xi^{-1} = \xi_{\alpha}^{-1} - \xi_{\beta}^{-1}$
and $r=N-p\gg 1$. The density approaches its bulk value $\rho_{bulk}
= \alpha$ from above. The current is given by $j=\alpha(1-\alpha)$
and therefore according to (\ref{20a}) $\langle \tau_N \rangle =
\alpha(1-\alpha)/\beta > \alpha=\langle \tau_1 \rangle$.
\vonecm \\
\udl{\bf Low density phase \AII:}
\vspace*{1cm} \\
The profile in this regime ($\beta > 1/2, \alpha < 1/2$)
is obtained from (\ref{33}) by exchanging
$\alpha$ and $\beta$ and substituting $p$ by $r=N-p$. This is a
result of the particle-hole symmetry of the model. In the same way
one obtains the profile on the phase transition lines from \AI to
\AII
and from \AII to C out of
the profiles on the phase transition lines from
\BI to \BII and \BII to C respectively.
\vspace*{1cm} \\
\udl{\bf Coexistence line:}
\vspace*{1cm} \\
If $\alpha = \beta < 1/2$
both $\xi_{\alpha}$ and $\xi_{\beta}$ are finite, but
since $\xi_{\alpha} = \xi_{\beta}$, one gets $\xi^{-1} = 0$.
As a result one finds a linear profile with a positive slope
\be\label{30}
t_N(p) = (1-2\alpha) / N \htwo .
\ee
The current is given by $j=\alpha(1-\alpha)$ and one has $\langle
\tau_1 \rangle = \alpha$ and $\langle \tau_N \rangle = 1 - \alpha$.

\section{Discussion of the phase diagram}
\setcounter{equation}{0}

We turn now to discuss the various phases and the transitions between
them on a more physical, intuitive basis. First we consider the case
$\beta=1$. This situation corresponds to connecting the system to a
reservoir of fixed density $\rho_0=\alpha$ at the origin, and another
"reservoir" with $\rho_{N+1} = 1- \beta =0$ at the boundary. We will
consider the limit $N \rightarrow \infty$, and ask what are the
possible \SS\ density profiles that the system can have, and which
interpolate between the two limiting values $\rho_0$ and $\rho_{N+1}$.

Let us start with $\alpha < 1/2$, and try a density profile
$(a)$ that approaches (for $1 \ll x \ll N$)
a constant bulk value $\rho < \alpha$, before it decays to
$\rho_{N+1}=0$. We now show that such a profile {\em cannot} be a \SS .
To see this, note that in a bulk region with constant density there
are no correlations (the steady state factorizes into a product
measure) and therefore the current in the bulk is given by
$j=\rho (1 - \rho)$; whereas at the origin it is
$j_0=\alpha (1-\rho_1)$, where $\rho_1$, the density at $x=1$,
satisfies  $1/2 > \alpha \geq \rho_1 \geq \rho$. If we can show
that $j_0 > j$, particles accumulate between the origin and the
bulk, and hence the density is not stationary. Clearly, for $\rho_1=\rho$
 we have $j_0=\alpha (1-\rho ) > \rho (1 - \rho)=j$  since
$\alpha > \rho$. On the other hand, for $\rho_1 = \alpha$
we have $j_0 = \alpha (1 - \alpha) > \rho (1- \rho)=j$ for $1/2 >
\alpha > \rho$.  Hence $j_0 > j$ at the two endpoints of the
interval $[ \rho , \alpha ]$ to which $\rho_1$ is limited; and since
$j_0$ is a linear function of $\rho_1$, we must have $j_0 > j$ for the
entire interval.

A different possible \SS\ profile $(b)$ is one with $\alpha < \rho
< 1/2$. Here we can show that $j_0=\alpha (1- \alpha) < \rho
(1-\rho) = j$: Under the present assumptions
the density first interpolates between $\alpha$ and $\rho$, and hence
$\alpha < \rho_1 < \rho$, and as before, the relationship
we wish to prove holds at both endpoints of this interval.
If this holds, however, more particles leave the bulk
than enter it, and $\rho$ must decrease. Thus also $(b)$ cannot be
a \SS .

The last possibility of the kind considered, $(c)$, has $\rho > 1/2
> \alpha$; we will return to this case later and show that for
the presently used values of $\alpha$ and $\beta$ it cannot be a \SS\
profile either. The only remaining situation is the one in which
$\rho = \alpha$. Then, obviously, $j_0=\alpha(1-\alpha)=j$. Hence the
bulk \SS\ density must equal that of the reservoir.

The assumption $\alpha < 1/2$ was crucial for our proof of this
fact, which is no longer true if $\alpha > 1/2$. In that case
the bulk density is, independently of $\alpha$, given by $\rho=1/2$.
To see this, we again assume all other possible values for the bulk
\SS density, and rule out every other scenario. Let us start by assuming
a decay to a bulk value $\rho < 1/2$; since, supposedly, we are in a
\SS, we can choose some point $i$ at which $1/2 > \rho_i >
\rho$ as a new initial point of fixed density $\alpha^\prime = \rho_i$;
$\alpha^\prime$ plays now the role of $\alpha < 1/2$ of the
previously discussed situation $(a^\prime)$,
which, as we have shown, cannot
be a \SS. Another possibility $(b^\prime)$ has $\alpha > \rho
> 1/2$. In this case we recall that near $N$ the density profile
must go from $\rho$ to $\rho_{N+1}=0$. To rule this out, we view the
site $N+1$ as a reservoir of {\em holes} of fixed density  1. The bulk
with $\rho > 1/2$ corresponds to hole density $\rho_h=1-\rho
< 1/2$; holes move to the left, and if we exchange the roles of
holes and particles, this situation becomes precisely the case
$(a^\prime)$ discussed above. Hence $(b^\prime)$ is not possible either.
\footnote{Note that the situation $(c)$ of the $\alpha < 1/2$ case,
to which we promised to return, also requires that $\rho$ goes from
$\rho > 1/2$ to $\rho_{N+1}=0$, and therefore is ruled out in the
same way as $(b^\prime)$.}

We have just shown that for $\alpha > 1/2$ no \SS\ is possible
with either bulk density $\rho < 1/2$ or $\rho > 1/2$; hence
the only possibility left is $\rho_{bulk}=1/2$. That is, for $\alpha
> 1/2$ the bulk density is that one which supports the maximal
current, irrespective of $\alpha$, the density of the reservoir. This
explains the transition observed at $\alpha = 1/2$, from a low density
phase with $\rho_{bulk}=\alpha$ to the maximal current phase, in which
$\rho_{bulk}=1/2$.

For the sake of convenience we limited the previous discussion to the
$\beta=1$ line. We now show that the transition survives when we move
off this line. Of the cases discussed above, $(a), (b)$ and $(a^\prime)$
were ruled out with no mention of the fact that $\beta=1$. In case
($b^\prime$) (and its equivalent, case ($c$)), we used a particle-hole
symmetry to map the situation onto case $(a^\prime)$. Since there
we had $\alpha > 1/2$, the same argument goes through for
holes if $\beta > 1/2$; hence the same considerations
give rise to the same phases as were obtained on the $\beta=1$
line. This completes the picture for the regions $1/2 < \beta
\leq 1$, and by particle-hole symmetry, for $1/2 < \alpha \leq 1$
as well.

Note that in the low density phase \AII with $\rho_{bulk}=\alpha$
the slope of the density
profile changes sign on the curve $\alpha=1-\beta$. This can be
understood as follows. The probability
that a particles moves in the bulk (its average velocity) is
$v = 1-\alpha$, while the probability that it moves at the boundary
is $v_N = \beta$. If $\beta > 1-\alpha$ then $v_N > v$ and the
system becomes depleted near the boundary because the current $j=\rho v$
is conserved. This corresponds the negative slope of the profile
in this regime. On the other hand, if $\beta < 1-\alpha$ one has
$v_N < v$ and particles pile up. This leads to a positive slope.
In the high density phase \BII one finds the same result when
compairing the velocity $v_0=1-\alpha$ at the origin with the
bulk velocity $v=\beta$.

Next, we discuss the low density phase \AI,
the high density phase \BI in the region $0 < \alpha, \beta
< 1/2$ and the transition between them.
The bulk values in both phases
and the slope of the profile can be derived in the same way as for
the phases \AII and \BII . Note that in both phases one has
$v_0 = 1-\alpha > \beta = v_N$ and therefore the slope is always
positive (particles pile up). The current is given by
\be\label{x1}
j = \min{(\alpha(1-\alpha),\beta(1-\beta))} \htwo .
\ee
In order to understand the shape of the
profile in phases \AI and \BI we assume that it is built up by a
superposition of profiles with a constant density $\alpha$ up
to some point $x_0$, followed by
constant density $1-\beta$. We call this sudden increase of the
average density a domain wall \footnote{We assume the width of this
domain wall to be very small compared to the size of the system.}
since it separates a region of high density $1-\beta$ from a region of
low density $\alpha$. The picture
we have in mind for this scenario is that particles injected with
rate $\alpha$ at the origin move with constant average velocity
$1-\alpha>1/2$ until they hit the domain wall where they get stuck
and continue to move only with velocity $\beta<1/2$. This region
of high density is caused by the blockage introduced through the
connection to the reservoir of density $1-\beta$ at the boundary.
Such a scenario is plausible, since constant densities $\alpha<1/2$
starting from the origin and $1-\beta > 1/2$ connected to the
boundary are both stable situations of the system as discussed above.
The probability $p(x)\propto\exp{(-x/\xi)}$ of finding this
domain wall at position $x$
is determined by the length scale $\xi$ given by $\xi^{-1}=
\xi_{\alpha}^{-1} - \xi_{\beta}^{-1}$ (see (\ref{31}) and (\ref{29})).
If $\alpha<\beta$ (low density phase) then particles are absorbed
with a higher probability than they are injected and the probability
of finding the domain wall decreases
exponentially with increasing distance $r=N-x$ from the boundary.
On the other hand, in the case where $\alpha>\beta$ (high density
phase) the situation is reversed and $p(x)$
decreases with increasing distance from the origin.
Averaging over all such profiles with the weight $p(x)$ leads to
the observed exponential decay to the respective bulk value.
This picture provides also a natural explanation of the linear profile
on the transition line $\alpha=\beta$ where the absorption and
injection probabilities are equal. Here $\xi$ diverges and
the probability of finding the domain wall at $x$ is independent of $x$.
Averaging over step functions with an equal weight for every
position of the step clearly gives a linear profile.
It is worth noting that
the mean field calculation done in \cite{1} gives a correct
description of the phases \AI and \BI, but it singles out
the constituent step function with the domain wall located in the center
as the profile on the phase transition line.
We should mention that this analysis, in particular the use of the
domain wall picture for a description the two phases and the phase
transition line, is based on our studies of a similar exclusion
process with \OBC\ but parallel dynamics \cite{G2}. For this model
we found phases of type \AI and \BI and a  phase transition separating
them as in the system with sequential dynamics studied here.
A careful study of the equal time correlation functions leads to our
interpretation in terms of domain walls. As the transition lines
to the the phases \AII or \BII are approached, this picture becomes
invalid.

Finally we briefly discuss the phase transition from the
high density phase \BI to the high density phase \BII .
On approaching $\alpha=1/2$ (but $\beta<1/2$) the length scale
$\xi_{\alpha}$ diverges while $\xi_{\beta}$ remains finite.
As a result neither the bulk density nor the way how
the bulk density is approached depends on $\alpha$ (except
for the trivial fact that the density at the origin and
consequently the amplitude of the derivative of the profile
depend on $\alpha$). The decay to the bulk density $\rho_{bulk}=1-\beta$
is determined by $\xi_{\beta}$ alone. Similarly the current
does not depend on $\alpha$, being $j=\beta(1-\beta)$.
A description of phase \BII also in terms of constituent profiles
is appealing at first sight, but it is less convincing since
a constant profile of density $\alpha > 1/2$ at the origin
is {\em not} a stable situation. In order to get a more intuitive
insight regarding this phase transition, we consider again the transition
from the low density phase \AII to the maximal current phase on the
line $\alpha=1/2$ but $\beta>1/2$. In the maximal current phase $C$ the
bulk density and the way how it is approached does not depend on $\alpha$
whereas in the low density phase \AII $\alpha$ does determine the bulk
density and how the profile decays to it.
This is obviously due to the fact that if $\alpha > 1/2$ the particles
close to the origin block each other rather than flowing away. As a
result, the information corresponding to a change in the injection
rate does not penetrate into the system. Clearly this description
of the effect of $\alpha$ increasing beyond 1/2
on the transition from the low density phase \AI to the maximal current
phase does not depend on the absorption at the boundary and is
therefore also applicable to the transition from phase \BI to phase
\BII .\footnote{
As a result of the particle-hole symmetry of the problem, the discussion
of the transition from the low density phase \AI to the low density
phase \AII proceeds along analogous lines.}

We conclude that the phase transitions to the maximal current phase
from the phase \AII (or \BII) is of the same nature as the
phase transition from \BI to \BII (or from \AI to \AII).
These transitions
are caused by reaching the maximal transport capacity of the
system at the origin (or boundary) and result in the divergence
of the corresponding length scale determining the shape of the
profile. Note that in our explanation it was necessary to take into
account
local correlations rather than only the current. This is the reason
why these phase transitions are not found in the mean field calculation.

As opposed to these transitions,
the one at $\alpha=\beta$ that takes the system
from the low density phase \AI to the
high density phase \BI is caused by the building up of domain walls.
Such a wall is generated by the inhomogenity forced on the system
by being connected to two reservoirs of different densities.
At the transition line the wall can be anywhere with equal probability.
The ``coexistence" of low and high density regions is known to occur
also in other systems with such an inhomogenity \cite{JL,G1,G2,KM}.

{\em Note added:} After completion of this work we received a
preprint by B. Derrida, M.R. Evans, V. Hakim and V. Pasquier, who
solved the same problem by a different method.

\section*{Acknowledgments}
We thank B. Derrida and D. Mukamel for useful discussions. This
research was partially supported by the Deutsche Forschungsgemeinschaft
and the US-Israel Binational Science Foundation.
\appendix
\renewcommand{\theequation}{A.\arabic{equation}}
\renewcommand{\thesection}{\mbox{Appendix }}
\setcounter{equation}{0}
\section{Some useful identities for the $G$-function}

The function $G_{N,K}^M(x)$ was defined in eq. (\ref{16}) as
\be
G_{N,K}^{M}(x) = \sum_{r=0}^{M-1} b_{N,K}(r) x^r \hspace{2cm} (N\geq 1)
\ee
with
\be
b_{N,K}(r) = \left( \ba{c} K-2+r \\ K-2 \ea \right) -
             \left( \ba{c} K-2+r \\  N  \ea \right) \hspace{2mm} .
\ee
Furthermore we defined $b_{0,1}(0)=G_{0,0}^1=G_{0,1}^1=1$.
Using the symmetries of the coefficients $b_{N,K}(r)$ it is easy to
prove the following recursion relations:
\be\label{A1}
\ba{lclr}
G_{N,K}^M(x) & = & G_{K-1,K}^M(x) + x G_{N,K+1}^{M-1}(x)  \hspace{2cm} &
(2 \leq K \leq N+1) \vfour \\
G_{N,K}^N(x) & = & G_{N,K-1}^N(x) + x G_{N-1,K}^{N-1}(x)  \hspace{2cm} &
(2 \leq K \leq N+1) \vfour \\
G_{N,N}^M(x) & = & G_{N-1,N-1}^M(x) + x G_{N,N}^{M-1}(x)  \hspace{2cm} &
(N \geq 2) \htwo .
\ea\ee
One also finds
\be\label{A2}
\ba{cccccccc}
G_{N,N}^N(x) & = & G_{N,N+1}^N(x) & = & G_{N,N}^{N+1}(x) & = &
G_{N,N+1}^{N+1}(x) & (N \geq 1) \vfour \\
G_{N,K}^0(x) & = & G_{N,N+2}^M(x) & = & 0 & & & \vfour \\
G_{N,K}^1(x) & = & G_{N,1}^M(x) & = & 1 & & & (K,M \leq N+1) \htwo .
\ea\ee
Eqs. (\ref{A1}) and (\ref{A2}) are necessary to prove that the
function $Y_{N,K}(\alpha,\beta)$ and $X_{N,K}^p(\alpha,\beta)$ satisfy
the recursion relations and initial conditions (\ref{12}) - (\ref{15}).

$G_{N,K}^M(x)$ can be expressed in terms of incomplete $\beta$-functions:
\be\label{A4}
(1-x)^{K-1} G_{N,K}^M(x) = I_{1-x}(K-1,M)
- \left(\frac{x}{1-x}\right)^{N-K+2} I_{1-x}(N+1,M-N+K-2)
\ee
(This is a direct consequence of the definition of $I_{x}(P,Q)$.)
{}From this expression one obtains
\be\label{A5}
(1-x)^{N+1} G_{N,K}^M(x) - x^M G_{M-1,M-N+K-1}^{N+1}(1-x) =
(1-x)^{N-K+2} - x^{N-K+2}
\ee
and by setting $M=N$ and $M=N+1$
\be\label{A3}
G_{N,K}^{N+2}(x) = (1-x) G_{N+1,K+1}^{N+1}(x) \htwocm
(2 \leq K \leq N+1) \htwo .
\ee
{}From this relation one can see that
relations (\ref{A1}) are consistent for $M=N=K$.

In the expression (\ref{22}) for the density profile only the
function $G_{L,L}^L(x)$ appears. From eq. (\ref{A3}) one finds
\be\label{A6}
G_{L,L}^L(x) = (1-x) G_{L+1,L+1}^{L+1}(x) + c_L \, x^{L+1}
\ee
with
\be\label{A7}
c_L = - b_{L,L}(L+1) = \frac{(2L)!}{L!(L+1)!}     \htwo .
\ee
Defining $F_L(x)=x^{(-L-1)}G_{L,L}^L(x)$ one gets
\be\label{A8}
F_L(x) = \frac{1-x}{\left(x(1-x)\right)^{L+1}} -
         \sum_{k=0}^{L-1} c_k \left(x(1-x)\right)^{k-L} \htwo .
\ee
Now we can reexpress $F_L(x)$ for $x\neq 1/2$
in terms of a hypergeometric function:
\be\label{A9}
\ba{rcl}
F_L(x) & = & \disp \frac{(1-2x)\Theta(1-2x)}{\left(x(1-x)\right)^{L+1}}
            \vfour \\
& &\disp + \, \frac{c_L}{(1-2x)^2}
F(1,\mbox{\small $\frac{3}{2}$};L+2;
\mbox{\small $\frac{-4x(1-x)}{(1-2x)^2}$ }) \vfour \\
 & = & F_L^{(1)}(x) + F_L^{(2)}(x)
\ea\ee
Here $\Theta(z)$ is the step function
\be\label{A10}
\Theta(z) = \left\{ \ba{ll} 1 \honecm & z>0 \vfour \\
                            0         & z<0 \ea \right. \htwo .
\ee
For large $L$ the hypergeometric function $F(1,\smfrac{3}{2};L+2;z)$
reduces to
\be
F(1,\smfrac{3}{2};L+2;z) = 1 + O(L^{-1}) \htwo .
\ee
Special exact values of $F_L(x)$ are
\be\label{A11}
F_L(\half) = 2 \BIN{2L}{L} = \frac{2}{\sqrt{\pi}} \frac{4^L}{L^{1/2}}
                             \left(1 + O(L^{-1}) \right)
\ee
and
\be\label{A12}
F_L(1) = c_L = \frac{4^L}{\sqrt{2\pi} L^{3/2}}
                             \left(1 + O(L^{-1}) \right) \htwo .
\ee

\newpage
\bibliographystyle{unsrt}

\end{document}